\documentclass[a4paper,11pt]{article}
\usepackage{pos}

\title{RI${}'$/SMOM renormalization for lattice QCD:
  bilinear and three-quark operators}

\author*[a]{B.~A.~Kniehl}
\author[b]{O.~L.~Veretin}

\affiliation[a]{II~Institut f\"ur Theoretische Physik, Universit\"at Hamburg,\\
  Luruper Chaussee 149, 22761 Hamburg, Germany}
\affiliation[b]{Institut f\"ur Theoretische Physik, Universit\"at Regensburg,\\
  Universit\"atsstrasse 31, 93040 Regensburg, Germany}

\emailAdd{kniehl@desy.de}
\emailAdd{veretin@mail.desy.de}

\abstract{
We review perturbative matching between the regularization-invariant
symmetric MOM (RI${}'$/SMOM) and $\overline{\rm MS}$ schemes at the symmetric
subtraction point, which is relevant for lattice QCD simulations.
For \emph{bilinear} operators we summarize three-loop conversion factors for
local quark currents and for the $n=2,3$ twist-two moments of structure
functions.
For \emph{three-quark} operators we summarize two-loop RI${}'$/SMOM matching
for $N=0$ baryonic operators and three-loop anomalous dimensions together
with two-loop matching for the $N=1$ Mellin moment, enabling improved lattice
studies of baryon distribution amplitudes.
All numerical results quoted here are given in Landau gauge.
}

\FullConference{Loops and Legs in Quantum Field Theory (LL2026)\\
12-17, April, 2026\\
Bayreuth, Germany\\}

\newcommand{\pslash}[1]{#1\!\!\!/\,\,}
\newcommand{\slashed}[1]{#1\!\!\!/\,\,}

\begin{document}
\maketitle

\section{Introduction}

The lattice formulation of quantum chromodynamics (QCD) provides access to
long-distance operator matrix elements from first principles using Monte Carlo
methods.
Many important physical observables---quark masses, moments of parton
distributions, baryon distribution amplitudes (DAs), and others---are related
to matrix elements of local composite operators.
Such matrix elements are intrinsically nonperturbative and cannot be obtained
from continuum perturbation theory alone.
However, to compare lattice data with high-energy phenomenology and with
results from QCD sum rules, it is necessary to express them in the
$\overline{\rm MS}$ renormalization scheme, which is the worldwide standard in
perturbative QCD.

On the lattice, nonperturbative renormalization is commonly performed in a
momentum-subtraction scheme~\cite{Martinelli:1994ty}.
The original RI/MOM and RI${}'$/MOM prescriptions subtract Green functions at
vanishing operator momentum, which can generate additional sensitivity to
short-distance effects.
The regularization-invariant symmetric MOM (RI/SMOM, RI${}'$/SMOM) scheme
proposed in Ref.~\cite{Sturm:2009kb} performs the subtraction at a symmetric
Euclidean point $p_i^2=-\mu^2$ where none of the external momenta is exceptional:
This improves the infrared behaviour compared with schemes where
one of the momenta is zero, as for example in the pseudoscalar channel where pion-pole
contributions can induce power-suppressed but numerically relevant
condensate effects of order $O(\Lambda^2_{\rm QCD}/\mu^2)$.

The conversion from RI${}'$/SMOM to $\overline{\rm MS}$ can be evaluated
perturbatively as a power series in $a=\alpha_s(\mu)/(4\pi)$ at a scale
$\mu$ of a few GeV, matching lattice simulations with continuum perturbation theory.
For bilinear quark operators, one- and two-loop conversion factors are known
from Refs.~\cite{Sturm:2009kb,Gracey:2010ci,Gorbahn:2010bf,Almeida:2010ns,Gracey:2011fb,Gracey:2011zn,Gracey:2011zg}.
For three-quark operators relevant to baryon DAs, one-loop matching and
two-loop anomalous dimensions have been studied in Refs.~\cite{Gruber:2017ozo,Bali:2015ykx,RQCD:2019hps,Bali:2024oxg}.

In this review we collect our own contributions to this field, spanning
projects on bilinear and three-quark operators in RI${}'$/SMOM kinematics.
The paper is organized as follows.
Section~\ref{sec:bilinear} treats bilinear operators: local currents and
twist-two moments.
Section~\ref{sec:threequark} treats three-quark operators.
Section~\ref{sec:conclusions} contains our conclusions.

\begin{figure}[h]
\centerline{
   \includegraphics[width=0.3\textwidth]{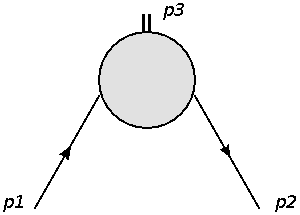}
   \hspace*{20mm}
   \includegraphics[width=0.3\textwidth]{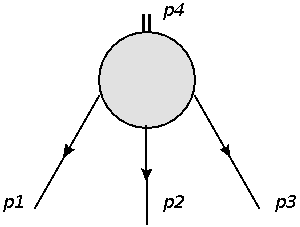}
}
\hspace*{40mm} (a) \hspace*{60mm} (b)
\caption{
  Matrix elements of a bilinear quark operator (a) and three-quark operator in momentum space. 
  Black box denotes operator. Solid lines denote external fermions. All momenta are incoming.
  }
\end{figure}

\section{Bilinear operators}
\label{sec:bilinear}

For a bilinear quark operator we start from the Minkowski-space correlator
\begin{align}
  \int dx\,dy \,e^{-iq\!\cdot\!x-ip\!\cdot\!y} \,\langle \psi_{\xi,i}(x) \,O_{\mu\dots\nu}(0)\,
  \bar{\psi}_{\zeta,j}(y) \rangle =
  \delta_{ij} \, S_{\xi\xi'}(-q)
  \Lambda_{\xi'\zeta'}(p,q) S_{\zeta'\zeta}(p)\,,
\label{eq:matrix_element}
\end{align}
where $O_{\mu\dots\nu}$ denotes operator, $\xi,\zeta$ are spinor indices, $i,j$ are colour indices, $S(p)$ is the
quark propagator, and $\Lambda(p,q)$ is the amputated Green function.
Such a matrix element is depicted in Fig.~1~(a) with $p_1=p$ and $p_2=q$.

For the subtraction we choose the symmetric kinematic point
\begin{align}
  p^2 = q^2 = (p+q)^2 = -\mu^2 \,, \qquad p\!\cdot\!q = \tfrac12\mu^2 \,.
\label{eq:kinematics}
\end{align}
This renormalization condition fixes the finite parts of the
form factors appearing in the tensor decomposition of $\Lambda$.
The conversion factor to $\overline{\rm MS}$ is then obtained as the ratio of
the perturbatively computed $\overline{\rm MS}$ and RI${}'$/SMOM renormalized
vertices with the same kinematics.

\subsection{Local currents}
\label{sec:bilinear-twist}

First we consider the scalar, vector, and tensor nonsinglet bilinear operators
\begin{align}
  J^S = \bar{\psi}\psi\,, \qquad
  J^V_\mu = \bar{\psi}\gamma_\mu\psi\,, \qquad
  J^T_{\mu\nu} = \bar{\psi}\sigma_{\mu\nu}\psi\,,
\label{bilinear_currents}
\end{align}
with $\sigma_{\mu\nu}=\frac{i}{2}[\gamma_\mu,\gamma_\nu]$.

Following Ref.~\cite{Gracey:2010ci} we decompose the amputated Green functions
into scalar form factors $F^{S,V,T}_j$ and tensor structures built from
$p$, $q$, and Dirac matrices:
\begin{align}
\label{eq:LambdaS}
  \Lambda^S &= \Gamma_0 \, F^S_1 + \mu^{-2}\Gamma_{2,pq} \, F^S_2\,, \\
  \Lambda^V_\mu &= \gamma_\mu F^V_1
      + \mu^{-2}\pslash{p}\, p_\mu F^V_2
      + \mu^{-2}\pslash{q}\, p_\mu F^V_3
      + \mu^{-2}\pslash{p}\, q_\mu F^V_4 \nonumber\\
     &\quad + \mu^{-2}\pslash{q}\, q_\mu F^V_5
      + \mu^{-2}\Gamma_{3,\mu pq} \, F^V_6 \,, \\
  \Lambda^T_{\mu\nu} &= \Gamma_{2,\mu\nu} F^T_1
      + \mu^{-2}(p_\mu q_\nu - p_\nu q_\mu) F^T_2
      + \cdots + \mu^{-4}\Gamma_{4,\mu\nu pq} F^T_8 \,,
\label{eq:LambdaT}
\end{align}
where $\Gamma_n$ denotes antisymmetrized products of $n$ $\gamma$-matrices with
$1/n!$ normalization.
In total there are 2 scalar, 6 vector, and 8 tensor form factors.

To evaluate the form factors we follow the standard two-step procedure.
First, Feynman integrals are reduced by integration-by-parts (IBP)
identities~\cite{Chetyrkin:1981qh} using FIRE~\cite{Smirnov:2014hma} to a
small set of master integrals.
At the symmetric kinematic point additional relations further reduce the
integral basis.
For bilinear operators at three loops the final result involves 2 one-loop,
8 two-loop, and 60 three-loop master integrals~\cite{Kniehl:2020sgo}.
Because some expansion coefficients $c_j(d)$ in
$A=\sum_j c_j(d)\,{\cal M}_j$ carry poles up to $1/\varepsilon^5$ while master
integrals have at most $1/\varepsilon^3$, an $\varepsilon$-finite master
integral basis is essential for numerical stability~\cite{Chetyrkin:2006dh}.

Master integrals are evaluated numerically by sector decomposition and Monte
Carlo integration using FIESTA~\cite{Smirnov:2015mct} and CUBA~\cite{Hahn:2004fe},
achieving relative accuracy $\sim10^{-6}$ for individual integrals.
Large cancellations in the sum over master integrals degrade the final
accuracy to $\sim10^{-4}$--$10^{-5}$, which is nevertheless sufficient for
lattice matching.

The RI${}'$/SMOM to $\overline{\rm MS}$ conversion factors for all 
currents (\ref{bilinear_currents}) have been computed through three loops~\cite{Kniehl:2020sgo}.
At one and two loops we find full agreement with
Refs.~\cite{Sturm:2009kb,Gracey:2010ci,Gracey:2011fb} after resolving a sign
convention for the momentum assignment.
The three-loop scalar result has been confirmed analytically in
Ref.~\cite{Bednyakov:2020ugu}.

In particular, for $SU(3)$ with $C_F=4/3$, $C_A=3$ we find
\begin{align}
  F^S_1 &= 1 + a \big( 0.6455188559544156 \big) \nonumber\\
    &\quad + a^2 \big( 48.48858821140752 - 6.346872802912313\, n_f \big) \nonumber\\
    &\quad + a^3 \big( 2396.16(7) - 417.872(2)\, n_f + 8.6453(1)\, n_f^2 \big) \,.
\end{align}
This quantity can be interprested as a conversion factor for quark mass defined
in $\overline{\rm MS}$ and RI${}'$/SMOM schemes.
The three-loop corrections are comparable in size to the two-loop ones and
should be included in precision lattice analyses of quark masses and bilinear
matrix elements.
Complete results for all form factors in general $SU(N)$ are available in
machine-readable form~\cite{Kniehl:2020sgo}.

\subsection{Twist-two moments $n=2,3$}
\label{sec:bilinear-twist}

Twist-two Wilson operators contributing to the $n$th moment of nonsinglet
structure functions are symmetric traceless combinations
\begin{align}
\label{eq:Odef}
   {\cal S} \bar{\psi} \gamma_{\mu_1} D_{\mu_2} \dots D_{\mu_n} \psi \,,
\end{align}
where ${\cal S}$ denotes total symmetrization with trace subtraction.
For $n=2$ one defines left- and right-derivative operators
\begin{align}
{\cal S}O^{L}_{\mu\nu} = {\cal S}\bar{\psi} \gamma_\mu
    \overset{\leftarrow}{D_\nu} \psi \,, \qquad
{\cal S}O^{R}_{\mu\nu} = {\cal S}\bar{\psi} \gamma_\mu
     \overset{\rightarrow}{D_\nu} \psi \,,
\end{align}
from which operators with definite charge conjugation $C=\pm1$ are built.
The basis used in Refs.~\cite{Gracey:2011zn,Gracey:2011zg} differs by a
$2\times2$ transformation; our conventions were chosen to ensure agreement
with those references at two loops.

For $n=2$ the amputated vertex decomposes into ten form factors $F^L_j$ and
$F^R_j$.
For $n=3$ one has three derivative assignments ($LL$, $LR$, $RR$) and
correspondingly more form factors.
We computed all conversion factors through three loops at the symmetric point
\eqref{eq:kinematics}~\cite{Kniehl:2020nhw}.

\paragraph{$n=2$ results.}
For the leading form factor $F^L_1$ in $\overline{\rm MS}$ we obtain
\begin{align}
  F^L_1 &= a \big( 0.87497670537933942370 \big) \nonumber\\
    &\quad + a^2 \big( 18.69246420435249196 - 2.5121840774766979282\, n_f \big) \nonumber\\
    &\quad + a^3 \big( 767.149(2) - 147.4921(2)\, n_f + 3.99834(1)\, n_f^2 \big) \,,\\
  F^L_2 &= 0.5 + a \big( -1.6874417634483485593 \big) \nonumber\\
    &\quad + a^2 \big( -21.75488024301858658 + 2.394054755622383881\, n_f \big) \nonumber\\
    &\quad + a^3 \big( -624.064(4) + 123.5347(3)\, n_f - 2.81311(1)\, n_f^2 \big) \,, \\
  F^L_5 &= a \big( -1.3116507463261931407 \big) \nonumber\\
    &\quad + a^2 \big( -30.32774293571586111 + 2.4134870684309379409\, n_f \big) \nonumber\\
    &\quad + a^3 \big( -1281.20(1) + 213.0615(9)\, n_f - 5.33964(3)\, n_f^2 \big) \,.
\end{align}
The remaining $F^L_j$ and the $F^R_j$ (obtained by crossing symmetry) are
listed in Ref.~\cite{Kniehl:2020nhw}.
Agreement with Ref.~\cite{Gracey:2011zn} at two loops is found via
$F^L_j=-\frac{1}{2}\Sigma^{W_2}_{(j)}$ and
$F^L_j+F^R_j=-\frac{1}{2}\Sigma^{\partial W_2}_{(j)}$.

\paragraph{$n=3$ results.}
For the $n=3$ moment the operator basis contains $LL$, $LR$, and $RR$
components.
As an illustration, the first three $F^{LL}_j$ form factors at three loops are
\begin{align}
F^{LL}_1 &= a \big( 0.12809418462663994519 \big)
 + a^2 \big( 3.57396324725023741 - 0.3927663257641307273\, n_f \big) \nonumber\\
 &\quad + a^3 \big( 142.934(4) - 23.3744(5)\, n_f + 0.38322(1)\, n_f^2 \big) \,, \\
F^{LL}_2 &= a \big( 0.64814814814814814815 \big)
 + a^2 \big( 11.92146760129898963 - 1.6685976202307340604\, n_f \big) \nonumber\\
 &\quad + a^3 \big( 470.434(6) - 94.2894(8)\, n_f + 2.69132(1)\, n_f^2 \big) \,, \\
F^{LL}_3 &= \tfrac13 + a \big( - 1.5801847945918187101 \big) \nonumber\\
 &\quad + a^2 \big( - 23.39093305099714828 + 2.878059562968590479\, n_f \big) \nonumber\\
 &\quad + a^3 \big( - 784.543(3) + 157.3102(3)\, n_f - 4.15775(1)\, n_f^2 \big) \,.
\end{align}
As for the local currents, the three-loop corrections are numerically
comparable to the two-loop ones.
These conversion factors are ready for use in lattice determinations of low
moments of parton distributions and structure functions.

\section{Three-quark operators}
\label{sec:threequark}

\subsection{Physical motivation and operator basis}

Light-cone distribution amplitudes describe the partonic structure of baryons
in exclusive processes~\cite{Efremov:1978cu,Lepage:1980fj,Chernyak:1983ej}.
Their Mellin moments are related to matrix elements of local three-quark
operators~\cite{Braun:2000kw}.
Lattice QCD provides a first-principles route to these matrix
elements~\cite{Bali:2015ykx,Bali:2024oxg,RQCD:2019hps}, but requires
perturbative conversion from RI${}'$/SMOM to $\overline{\rm MS}$.

The non-local three-quark operator with open spinor indices reads
\begin{eqnarray}
\label{eq:nonlocal}
O_{\xi_1\xi_2\xi_3}&=&(\slashed{n}u^{c_1^\prime})_{\xi_1}(nx_1)[nx_1,nx_0]^{c_1^\prime c_1}
(\slashed{n}d^{c_2^\prime})_{\xi_2}(nx_2)[nx_2,nx_0]^{c_2^\prime c_2}\nonumber\\
&&{}\times(\slashed{n}s^{c_3^\prime})_{\xi_3}(nx_3)[nx_3,nx_0]^{c_3^\prime c_3}
\epsilon^{c_1c_2c_3}\,,
\end{eqnarray}
where gauge links render the operator gauge invariant and $\epsilon^{c_1c_2c_3}$
ensures baryon number and colour neutrality.
Local operators with $N$ covariant derivatives correspond to the $N$th Mellin
moment.
Of particular interest are the spin-$3/2$ and spin-$1/2$ projections
$O^{(\frac{3}{2},0)}_{+}$ and $O^{(1,\frac{1}{2})}_{-}$ introduced in
Ref.~\cite{Braun:1998id}.

Expanding (\ref{eq:nonlocal}) in operator product expansion on light-cone we
obtain as a basic object the amputated four-point function (see Fig.~1~(b)):
\begin{align}
&H_{\beta_1\beta_2\beta_3,\alpha_1\alpha_2\alpha_3}^{\mu_1\dots\nu_1\dots\sigma_1\dots}(p_1,p_2,p_3) =
    - \int d^4x_1 \, d^4x_2 \, d^4x_3
     e^{i(p_1\!\cdot\!x_1 + p_2\!\cdot\!x_2 + p_3\!\cdot\!x_3)}
     \epsilon^{a_1a_2a_3} \nonumber\\
  &\qquad {}\times
     \langle O^{\mu_1\dots\mu_m\nu_1\dots\nu_n\sigma_1\dots\sigma_s}_{\beta_1\beta_2\beta_3}(0)\,
     \bar{u}^{a_1}_{\alpha'_1}(x_1) \bar{d}^{a_2}_{\alpha'_2}(x_2) \bar{s}^{a_3}_{\alpha'_3}(x_3)
     \rangle \nonumber\\
  &\qquad {}\times
     G_2^{-1}(p_1)_{\alpha'_1\alpha_1}
     G_2^{-1}(p_2)_{\alpha'_2\alpha_2}
     G_2^{-1}(p_3)_{\alpha'_3\alpha_3} \,,
\end{align}  
where the three-quark local operators are defined as
\begin{align}
\label{qqq_op}
&O^{\mu_1\dots\mu_m\nu_1\dots\nu_n\sigma_1\dots\sigma_s}_{\xi_1\xi_2\xi_3}(x) =
     \epsilon^{b_1b_2b_3} (D^{\mu_1}\dots D^{\mu_m}u^{b_1}_{\xi_1}) 
             (D^{\nu_1}\dots D^{\nu_n}d^{b_2}_{\xi_2})
             (D^{\sigma_1}\dots D^{\sigma_s}s^{b_2}_{\xi_3}) \,,
\end{align}  
where all quark fields are taken at the same space-time point $x$.
In (\ref{qqq_op}) the total number of derivatives $m+n+s=N$ defines the $N$th Mellin's moment of the DA.
We considered the cases of $N$ equals 0, 1 and 2.

In RI${}'$/SMOM kinematics for four external legs one adopts~\cite{Bali:2015ykx,Kniehl:2022ido}
\begin{align}
\label{eq:kin4}
p_1^2 = p_2^2 = p_3^2 = p_4^2 = -\mu^2 \,, \quad
p_1\!\cdot\! p_2 = p_1\!\cdot\! p_3 = \frac{1}{2}\mu^2\,, \quad
p_2\!\cdot\! p_3=0\,,
\end{align}
with $p_4=-(p_1+p_2+p_3)$ the momentum flowing into the operator.
All external legs carry a common virtuality, which is the prescription used in
recent lattice studies of baryon DAs~\cite{Bali:2024oxg,RQCD:2019hps}.

\subsection{Open spinor indices and evanescent operators}

Three-quark operators require special care in dimensional regularization.
In $d=4-2\varepsilon$ dimensions there exist infinitely many independent Dirac
tensor structures, while only finitely many survive at $d=4$.
Operators built from antisymmetric tensors of rank $n>4$ vanish in four
dimensions and are called \emph{evanescent}; they nevertheless mix with
physical operators under renormalization and yield finite contributions.
Following Ref.~\cite{Krankl:2011gch}, we renormalize three-quark operators
\emph{without} contracting spinor indices.
This guarantees that evanescent operators vanish in $d=4$ and systematically
avoids $\gamma_5$ ambiguities, at the price of a larger mixing matrix:
for $N=0$ one has $64$ components at one loop and $247$ at two loops in four
dimensions (Table~\ref{tab:N}).

\begin{table}[h]
\centering
\begin{tabular}{|c|c|c|c|}
 \hline
  \# of loops & 0 & 1 & 2 \\
 \hline
  $N$ (in $d$ dimensions) & 1 & 67 & 581 \\
 \hline
  $N$ (in 4 dimensions) & 1 & 64 & 247 \\
 \hline
\end{tabular}
\caption{Number of independent form factors for three-quark operators without
derivatives at different loop orders~\cite{Kniehl:2022ido}.}
\label{tab:N}
\end{table}

\subsection{Two-loop matching for $N=0$}
\label{sec:threequark-N0}

The amputated four-point function is decomposed as
\begin{equation}
\label{qqq_decomp}
  H_{\beta_1\beta_2\beta_3,\alpha_1\alpha_2\alpha_3}
  = \sum_{n=1}^{N} T_{n,\beta_1\beta_2\beta_3,\alpha_1\alpha_2\alpha_3}\,
    f_n(\mu^2) \,,
\end{equation}
where $T_n$ are spin tensor structures and $f_n$ scalar form factors.
At two loops in four dimensions there are $N=247$ independent structures
(Table~\ref{tab:N}).
Form factors are extracted by projecting with suitable tensors and renormalized
in $\overline{\rm MS}$ before setting $d=4$~\cite{Kniehl:2022ido}.

The simplest form factor $f_1$, corresponding to
$\Gamma_0\otimes\Gamma_0\otimes\Gamma_0$, reads in $R_\xi$ gauge
\begin{eqnarray}
  \frac{f_1}{f_{1,\mbox{Born}}} &=& 1 + a \big( 0.6204053307351691 + 0.5957023845688996\,\xi \big) \nonumber\\
  & &{}+ a^2 \big( 10.4515(4) + 3.5881(4)\xi + 1.4232(2)\xi^2 - 0.68933(2)\, n_f \big) \,,
\end{eqnarray}
with $f_{1,\mbox{Born}}=6$ and $a=\alpha_s/\pi$.
In Landau gauge ($\xi=0$) this gives the RI${}'$/SMOM to $\overline{\rm MS}$
matching for the lowest Mellin moment.
All 247 form factors are provided analytically in $R_\xi$ gauge and
numerically in Landau gauge in Ref.~\cite{Kniehl:2022ido}.

From the open-index results one can construct matrix elements of standard
baryonic currents, e.g.
\begin{eqnarray}
  (O_1)_\alpha &=& \epsilon^{ijk} u_\alpha^i \big[ (u^j)^T C d^k \big] \,,\\
  (O_2)_\alpha &=& \epsilon^{ijk} \gamma_5 u_\alpha^i \big[ (u^j)^T C \gamma_5 d^k \big] \,,
\end{eqnarray}
where $C$ is the charge-conjugation matrix.
For $O_1$ in Landau gauge we find at two loops
\begin{eqnarray}
O_1 &=& I\otimes I \big[1 + a(2.00280429) + a^2(26.486(1) - 2.79941(6)\, n_f)\big] \nonumber\\
& &{}+ \big(I\otimes \sigma_{p_1p_2} + I\otimes \sigma_{p_3p_1}\big)
    \big[ - a(0.52086828) + a^2(-10.85904(3) + 0.75237(3)\, n_f)\big] \nonumber\\
& &{}+ I\otimes \sigma_{p_2p_3} \big[ a(1.38629436) + a^2(34.345(1) - 2.4645(1)\, n_f)\big] \nonumber\\
& &{}+ \text{(further tensor structures)} \,,
\end{eqnarray}
demonstrating that the open-index formalism yields directly usable results for
lattice matching.

\subsection{Two-loop matching for $N=1$}
\label{sec:threequark-N1}

The amputated four-point functions for $N=1$ three-quark operators with open
spinor indices,
$H^{\mu,j}_{\chi_1\chi_2\chi_3,\xi_1\xi_2\xi_3}(p_1,p_2,p_3)$, is evaluated at
two loops in RI${}'$/SMOM kinematics~\eqref{eq:kin4}~\cite{Kniehl:2026qji}.
Here $j$ shows at which quark line the covariant derivative is acting.
Thus we have three operators, corresponding to $j=1,2,3$, which mix under renormalization.

At leading twist the decomposition (\ref{qqq_decomp}) modifies.
At one loop (two loops) the number of tensor structures in four dimensions
grows from $64$ ($247$) for $N=0$ to a larger set for $N=1$ due to the
additional Lorentz index from the covariant derivative. However in the leading twist
approximation the Lorentz and spinor indices factorize and we can write
\begin{equation}
\label{qqq_decomp1}
  H_{\beta_1\beta_2\beta_3,\alpha_1\alpha_2\alpha_3}^{\mu,j}
  = \sum_{n=1}^{N} p_j^\mu\,T_{n,\beta_1\beta_2\beta_3,\alpha_1\alpha_2\alpha_3}\,
    f_{n,j}(\mu^2) \,.
\end{equation}

The computational procedure follows Ref.~\cite{Kniehl:2022ido}:
renormalize the amplitude in $\overline{\rm MS}$, set $d=4$, and extract
form factors by projection.
Analytic results in terms of master integrals and numerical values in Landau
gauge are provided in Ref.~\cite{Kniehl:2026qji}.

Together with the three-loop anomalous dimensions, this allows one to extract
the first two Mellin moments ($N=0,1$) of baryon DAs from lattice QCD at
two-loop matching accuracy with three-loop scale evolution, as recently
demonstrated in Ref.~\cite{Bali:2024oxg}.

\subsection{Three-loop anomalous dimensions}
\label{sec:threequark-AD}

The $\overline{\rm MS}$ anomalous dimensions 
$\gamma_{(N)}=a\gamma^{(1)}_{(N)}+a^2\gamma^{(2)}_{(N)}+a^3\gamma^{(3)}_{(N)}+\dots$
for Mellin moments
$N=0,1,2$ have been computed through three loops~\cite{Kniehl:2026qji}.
For the $N=0$ operator with open spinor indices the result is particularly
compact:
\begin{eqnarray}
\label{eq:gamma0}
\gamma^{(1)}_{(0)} &=&-\tfrac{1}{3}\, \mathbb{C}_2\,,\\
\gamma^{(2)}_{(0)} &=& (70-4n_f) \mathbb{C}_0
  + \left( -\tfrac{49}{18} + \tfrac{n_f}{27} \right) \mathbb{C}_2
  + \tfrac{1}{9} \mathbb{C}_4 + \tfrac{1}{18} \mathbb{C}_{42}\,,\\
\gamma^{(3)}_{(0)} &=& \left( \tfrac{16492}{9} - \tfrac{1708}{9} n_f + \tfrac{20}{9} n_f^2 - \tfrac{434}{3} \zeta_3 \right) \mathbb{C}_0
  + \cdots \,,
\end{eqnarray}
where $\mathbb{C}_0=\Gamma_{000}$, $\mathbb{C}_2=\Gamma_{022}+\Gamma_{202}+\Gamma_{220}$, etc.
Here $\Gamma_{ijk}=\Gamma_i\otimes\Gamma_j\otimes\Gamma_k$ is the tensor product of three $\Gamma$-structures.
After rewriting evanescent structures via $d$-dimensional identities and
accounting for finite renormalizations, we find full agreement with
Refs.~\cite{Krankl:2011gch,Gracey:2012gx} for $N=0$.
Note that for $N=0$ the anomalous dimension is known now to four loops \cite{Gracey:2025jnh}.

For $N=1,2$ the anomalous dimensions involve the full set of $\Gamma_{ijk}$
structures; the three-loop results extend the two-loop $N=1$ analysis of
Ref.~\cite{Bali:2024oxg}.

For the spin-$3/2$ and spin-$1/2$ operators $O^{(\frac{3}{2},0)}_{+}$ and
$O^{(1,\frac{1}{2})}_{-}$, the anomalous dimensions through three loops are
also known~\cite{Kniehl:2026qji}.
In four dimensions these operators are eigenvectors of the leading-twist
renormalization matrix, which simplifies the extraction of physical
observables.

\section{Conclusions}
\label{sec:conclusions}

We reviewed perturbative RI${}'$/SMOM to $\overline{\rm MS}$ matching for
bilinear and three-quark operators at the symmetric subtraction point.

For \emph{bilinear} operators, three-loop conversion factors are now available
for local scalar, vector, and tensor currents~\cite{Kniehl:2020sgo} and for
twist-two moments $n=2,3$~\cite{Kniehl:2020nhw}.
In all cases the three-loop corrections are comparable in size to the
two-loop ones, so their inclusion reduces systematic uncertainties in lattice
extractions of quark masses and structure-function moments.

For \emph{three-quark} operators, two-loop RI${}'$/SMOM matching for the
$N=0$ Mellin moment~\cite{Kniehl:2022ido} and three-loop anomalous dimensions
for $N=0,1,2$ together with two-loop matching for $N=1$~\cite{Kniehl:2026qji}
advance the state of the art for baryon distribution amplitudes.
The open-spinor-index approach of Ref.~\cite{Krankl:2011gch} proved essential
for dealing with evanescent operators and $\gamma_5$ ambiguities.

Future work includes completing the two-loop matching for $N=2$ and further
refining the numerical evaluation of master integrals at three loops.
All results discussed here are available in machine-readable form in the
respective publications.

\acknowledgments

We are grateful to Vladimir M. Braun, Meinulf G\"ockeler, and Alexander N. Manashov for fruitful discussions.
O.L.V. is grateful to the University of Hamburg for the warm hospitality.
This work was supported in part by the German Research Foundation DFG through
Research Unit FOR~2926 ``Next Generation Perturbative QCD for Hadron Structure:
Preparing for the Electron-Ion Collider'' under Grant Nos.~KN~365/13-2 and 409651613.


\bibliographystyle{JHEP}
\bibliography{proc}

\end{document}